\documentclass[pra,showpacs,graphics,twocolumn,floatfix,mathbbm,a4paper,nofootinbib]{revtex4-2}

\usepackage{amsmath}
\usepackage{amsthm}
\usepackage{latexsym}
\usepackage{amsfonts}
\usepackage{amssymb}
\usepackage{bm,dsfont}
\usepackage{color}
\usepackage{graphicx}
\usepackage{enumerate}
\usepackage{subfigure}
\usepackage{multirow}

\usepackage[normalem]{ulem}
\usepackage{mathrsfs}
\usepackage{mathtools}

\usepackage{comment}

\usepackage{array}
\usepackage{hhline}

\newtheorem{theorem}{Theorem}

\theoremstyle{definition}

\definecolor{darkgreen}{rgb}{0,0.4,0.3}
\definecolor{gray}{rgb}{0.5,0.5,0.5}

\newcommand{\R}{\mathbb{R}} %real
\newcommand{\C}{\mathbb C}

\newcommand{\half}{\tfrac{1}{2}} %half
\newcommand{\mo}[1]{\left\vert #1 \right\vert} 
\newcommand{\floor}[1]{\left\lfloor{#1}\right\rfloor}

\newcommand{\lc}{\mathcal{L}(\C^2)} %linear operators on C2
\newcommand{\no}[1]{\left\|#1\right\|} %norm
\newcommand{\tr}[1]{{\rm tr}\left[#1\right]} %trace
\newcommand{\id}{\mathds{1}} %identity operator
\newcommand{\va}{{\bm a}} % a
\newcommand{\vb}{{\bm b}} % b
\newcommand{\vm}{{\bm m}} % m
\newcommand{\vn}{{\bm n}} % n
\newcommand{\vt}{{\bm t}} % t
\newcommand{\vx}{{\bm x}} % x
\newcommand{\vy}{{\bm y}} % y
\newcommand{\vi}{{\bm i}} % i
\newcommand{\vj}{{\bm j}} % j
\newcommand{\vsigma}{{\bm \sigma}}% sigma
\newcommand{\M}{\mathsf{M}}%generic observable
\newcommand{\Ms}{\M^\stan}
\newcommand{\Ma}{\M^\anti}
\newcommand{\en}{\mathcal{E}} %ensemble
\newcommand{\cpostt}{CPOST} %text CPOST
\newcommand{\cpostm}{{\scalebox{0.4}{\rm CPOST}}} %math CPOST
\newcommand{\anti}{{\rm an}}
\newcommand{\stan}{{\rm st}}
\newcommand{\sprob}{\mathbb{P}} %success probability
\newcommand{\Eg}{\sprob} %guessing average
\newcommand{\Es}{\sprob^{\,\stan}}
\newcommand{\Ea}{\sprob^{\,\anti}}
\newcommand{\Epost}{\sprob_{\cpostm}}
\newcommand{\Espost}{\Es_{\cpostm}}
\newcommand{\Eapost}{\Ea_{\cpostm}}

\newcommand{\cond}{\,|\,}

\begin{document}

\title[]{Anticipative measurements in hybrid quantum-classical computation}

\author{Teiko Heinosaari}
\affiliation{Quantum Algorithms and Software, VTT Technical Research Centre of Finland Ltd, Espoo, Finland}
\affiliation{Department of Physics and Astronomy, University of Turku, Turku 20014, Finland}
\author{Daniel Reitzner}
\affiliation{Quantum Algorithms and Software, VTT Technical Research Centre of Finland Ltd, Espoo, Finland}
\author{Alessandro Toigo}
\affiliation{Dipartimento di Matematica, Politecnico di Milano, Piazza Leonardo da Vinci 32, 20133 Milano, Italy}
\affiliation{Istituto Nazionale di Fisica Nucleare, Sezione di Milano, Via Celoria 16, 20133 Milano, Italy}

\begin{abstract}
Before the availability of large scale fault-tolerant quantum devices, one has to find ways to make the most of current noisy intermediate-scale quantum devices. One possibility is to seek smaller repetitive hybrid quantum-classical tasks with higher fidelity, rather than directly pursuing large complex tasks. We present an approach in this direction where the quantum computation is supplemented by a classical result. While the presence of the supplementary classical information helps alone, taking advantage of its anticipation also leads to a new type of quantum measurements, which we call anticipative. Anticipative quantum measurements lead to improved success rate over cases where we would use quantum measurements optimized without assuming the later arriving  supplementing information.
Importantly, in an anticipative quantum measurement the combination of the results from classical and quantum computations happens only in the end, without the need for feedback from the one to the other computation, a feature which hence allows for running both computations in parallel.
We demonstrate the method using an IBMQ device and show that it leads to an improved success rate even in a real noisy setting.
\end{abstract}

\maketitle

%%%%%%%%%%%%%%%%%%%%%%
\section{Introduction}\label{sec:intro}
%%%%%%%%%%%%%%%%%%%%%%

It is a trope of quantum computation to look for problems with a more advantageous complexity than their classical counterparts. With the advent of practical quantum computation, it became apparent that it is worthwhile to look also at problems where even a modest speedup can become useful. 
At this point, an additional aspect to be aware of is constituted by the limited resources of these devices.
In particular, current noisy intermediate-scale quantum (NISQ) devices have too many imperfections to be used for general universal quantum computation \cite{preskill2018quantum,bharti2022noisy}, hence approaches that are able to deal with these imperfections are seeked both
 in circuits \cite{zhang2014DD, edmunds2020qctrl, baum2021qctrl} and in measurements \cite{kwon2020hybrid,maciejewski2020mitigation,nachman2020unfolding,geller2020rigorous}.
A possible application of limited NISQ devices is quantum-classical hybrid computation \cite{bravyi2016trading,dunjko2018computational,angara2020hybrid,callison2022hybrid}, where the NISQ device has only a partial role, for example as a subroutine that is able to speed up a repeated computation task. 
One of the questions arising in these kinds of schemes is how to optimally combine classical and quantum computations so that their overall functioning is as good as possible.

\begin{figure}
\begin{center}
\includegraphics[width=6.5cm, height=6.5cm]{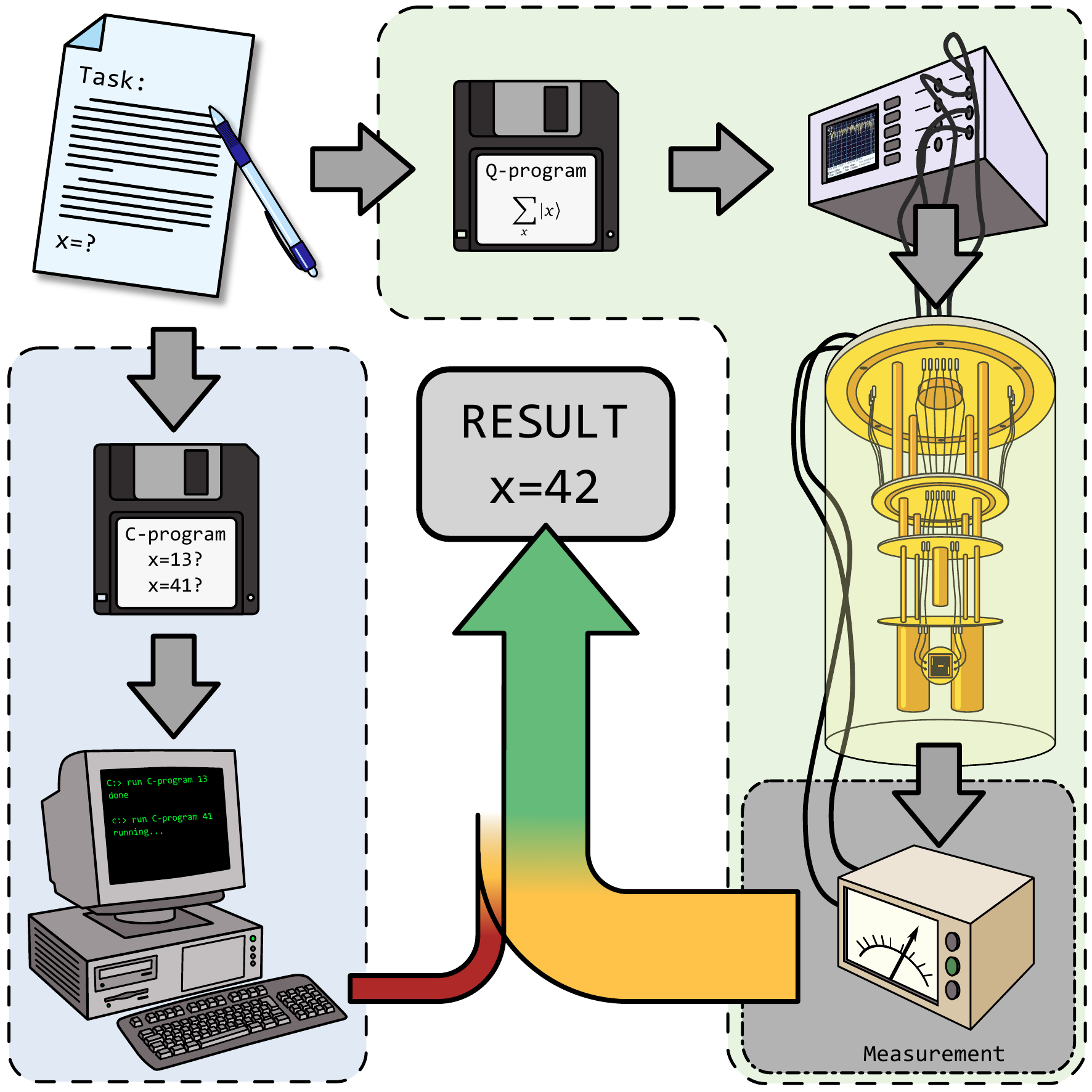}
\end{center}
\caption{We consider hybrid computation where classical and quantum computations are performed simultaneously. The outputs of the two computations are then combined to reduce the possible final answers and to provide a more precise output than from either of the two computations.\label{fig:hybrid_computation}}
\end{figure}

We are considering a class of tasks where one is required to compute the value of a function $g:X \to Y$ on any given input $x\in X$. 
Concrete examples of interest could be the following:
\begin{itemize}
\item $X = \{2,\ldots,n\}$, $Y=\{ 1,2, \dots, \floor{n/2} \}$, $g(x)=$ the largest proper divisor of $x$.
\item $X\,{=}$ the set of graphs with at most $n$ vertices, $Y=\{ 1,2, \dots, n \}$, $g(x)=$ the chromatic number of $x$.
\end{itemize}
In the current work, we do not concentrate on any particular problem, but rather on a method that is applicable to a variety of such tasks.
We focus on a scheme where the computation has two parallel parts: a classical part and a quantum part. 
The essential assumption in our investigation is that these computations are run in parallel, meaning that they take an input at the same time and the computations are carried out simultaneously. 
The combination of the computations and the final inference happens only after both parts have been completed (see Fig.~\ref{fig:hybrid_computation}).
The question is then if there is a way to optimize the two parts jointly that would lead to a better performance than individually optimizing each part.

The motivation for this parallel computing assumption is the following. Naturally, we want the total computing time to be minimal, which can be achieved by increasing the total success rate.
Considering current NISQ devices, the motivation may be even more practical. The available quantum computer may be small and less efficient than the computational problem would require, leading to intrinsically imperfect computation. A classical computer may provide a useful aid even if it cannot solve the problem efficiently alone. 
For instance, the classical computation can be used to rule out some wrong answers, which is much more easier than to solve the question in full. This partial information can still help the quantum computation part, and that is exactly the main idea of the current study.

We note that the method we are introducing does not prevent from having also sequential steps in the total computation process. Indeed, the method can be thought to be applied separately in all phases. We can even find further motivation in the sequential arrangement, since this kind of hybrid tasks uses classical computation between separate quantum computations and so, during each run of the quantum computation, classical computers are idle. In fact, the advantage of our approach is that it allows to make use of this ineffective time.

%%%%%%%%%%%%%%%%%%%%%%
\section{Parallel quantum-classical computation with anticipative measurements}
%%%%%%%%%%%%%%%%%%%%%%

%%%%%%%%%%%%%%%%%%%%%%
\subsection{Parallel quantum-classical computation}\label{sec:intro_anti}
%%%%%%%%%%%%%%%%%%%%%%

In our setting, the quantum computing part uses a $d$-level quantum system in the encoding of inputs and this size evidently limits the power of the quantum device. 
Each input $x$ is encoded into a quantum state $\varrho_x$. 
Then, the quantum system is acted on by a quantum process $\Phi$, hence the state is transformed into $\Phi(\varrho_x)$. 
The process can be a complicated array of quantum gates, it can use higher dimensional ancillary systems, and the produced states may possibly be not orthogonal.
Further, the transformed state $\Phi(\varrho_x)$ can be a state of a different quantum system having a different dimension than $d$. 
The system in the transformed state is measured by a quantum measurement $\M$ and an outcome $z$ is recorded.
The conditional probability to obtain $z$ is given by the Born rule
\begin{equation}
p(z \cond x) = \tr{ \Phi(\varrho_x) \M(z) } \, .
\end{equation}
In the most favorable case, one has $p(z \cond x)=\delta_{z,g(x)}$, i.e., the recorded outcome is the value $g(x)$ that we wanted to compute.
However, due to the limited size of the quantum computer and the noise involved in the process and in the final measurement, we expect $p(z\cond x)$ to be a probability distribution that has nonzero variance. 
The quantum computing part is therefore giving a probabilistic information about the correct answer for the value $g(x)$. 
 
In addition to the quantum computation, we assume that there is a classical computation part where the aim is to rule out one or more wrong answers. We denote by $S \subset Y$ the set of values that the classical computation determines to be wrong. For example, in NP-problems checking answers is a fast process that can be used to generate the set $S$.
Getting back to our earlier examples, if $g(x)$ is the largest proper divisor of an integer $x\geq 2$, the set $S$ may be obtained by performing divisibility tests for some choice of potential divisors.
Or, if $x$ is a graph and $g(x)$ is the chromatic number of $x$, $S$ could be the result of some random coloring heuristics carried out on $x$. 
The aim of the current investigation is not to identify specific problems where the classical computation can be explicitly described, and we actually let the set $S$ be any non-trivial subset of $Y$.

The limits on the computation time and the efficiency of the classical computer may imply that $S$ is a singleton set or has some other fixed size. 
In a typical application that we have in mind, $S$ is a small set compared to the size of $Y$.
The information in the form of $S$ is provided only after the quantum part has concluded and we will therefore call it the \emph{classical posterior information}, abbreviated to \cpostt{}. We further assume that the classical computation is error-free, so that $g(x) \notin S$. 
However, we allow it to be non-deterministic, and we denote the conditional probability to obtain $S$ given $x$ as $p(S \cond x)$. For instance, if we would use a quantum computer to speed up some NP-problem, alongside we could run some number of classical computations in which we may randomly choose potential answers $y_1, y_2,\ldots, y_m \in Y$, and either learn that one of our choices is the correct answer, or learn that $g(x) \not\in S=\{y_1, y_2,\ldots, y_m\}$. As the classical and quantum computations work in parallel and do not interact up to this point, we further require that the set $S$ of the excluded answers is independent of the measurement outcome $z$ when conditioned on the input $x$, that is, $p(S,z\cond x) = p(S \cond x)\,p(z\cond x)$.
  
The final step is then to combine the results from the two parts and make the final guess $y$, based on the outcome $z$ of the quantum computation and the set $S$ consisting of wrong values. In general, again, this guess is probabilistic and described by the conditional probability $p(y \cond S,z)$. Moreover, it depends on the input $x$ solely through the classical and quantum computations, meaning that $p(y \cond S,z) = p(y \cond S,z,x)$.
The efficiency of the full computation is characterized by its success probability, i.e., the probability of having $y=g(x)$ when the input $x$ is sampled from $X$. For simplicity, we will consider uniform sampling, although we remark that this is not a necessary requirement in our approach.

Let us now discuss different ways of utilizing posterior information. The setting provides a four-fold distinction between different scenarios.
Firstly, we can distinguish between the cases in which \cpostt{} is present or absent. 
A motivation for this distinction can be the case when \cpostt{} might improve the quality of the result from the quantum computation, but the result might be useful even without \cpostt{}. Secondly, and more importantly, we distinguish the cases where the quantum measurement is optimized without considering \cpostt{} (the resulting measurements will be called \emph{standard}) and where the quantum measurement is optimized by having \cpostt{} in mind (measurements of this kind will be called \emph{anticipative}).

The focus of this paper lies on the anticipative measurements with \cpostt{}. Our method thus optimizes the quantum and classical steps together instead of treating them separately, even if they are run independently in parallel.
In practice, this means that we adjust the measurement performed in the end of the quantum computation. 
The crux is to take into account that there will be \cpostt{}, in our case in the form of wrong answers, before we have to make the final decision, although we cannot know the specific wrong answers before we have to perform the measurement. 

It is not evident from this general description that the anticipative measurement method works and, in fact, it does not necessarily lead to a better success probability than the independently optimized parts. 
However, we demonstrate with examples that the anticipative measurement does give a benefit on a class of problems.
The main goal of the current investigation is to show that the anticipative method can be better than the standard method and therefore might be a valuable tool in hybrid quantum-classical computation.

To summarize, for a given task, we have four scenarios to which we assign the respective success probabilities $\Eg$.
\begin{enumerate}
\item $\Es_0$: standard quantum measurement without \cpostt{}. In this scenario, we use quantum computation only.
\item $\Espost$: standard quantum measurement with \cpostt{}.
This means that we run both quantum and classical computations, optimize them independently and combine in the end.
\item $\Eapost$: anticipative quantum measurement with \cpostt{}.
In this scenario, quantum measurement is optimized already having in mind the later arrival of classical information. This scenario is our main interest.
\item $\Ea_0$: anticipative quantum measurement without \cpostt{}.
This scenario is for comparison only. One can think of it as a scenario where classical computation breaks down and does not give any information even if we were waiting for it and hence chose anticipative measurement instead of the standard measurement.
\end{enumerate}
In general, we have
\begin{equation}
\Ea_0 \leq \Es_0 \leq \Espost \leq \Eapost\,.
\end{equation}
The first and last inequalities follow from the definitions. Indeed, the standard measurement means the optimal measurement for the problem without \cpostt{} and the anticipative measurement means the optimal measurement for the task with \cpostt{}. The middle inequality is true as posterior information cannot make the guessing probability worse if properly optimized (optimal solution must be at least as good as if one does not act based on \cpostt{}). 
Depending on the task in question, these inequalities may be equalities, which would mean that for that task posterior information or the anticipative method do not help.
Clearly, the anticipative method becomes interesting in tasks where $\Espost < \Eapost$. In Section \ref{sec:qubit}, we demonstrate that this is actually the case already in a simple class of tasks.

%%%%%%%%%%%%%%%%%%%%%%
\subsection{Mathematical framework for anticipative measurements}\label{sec:anticipativeframework}
%%%%%%%%%%%%%%%%%%%%%%

We now present the mathematical framework for anticipative measurements that is needed in the applications to concrete problems. 

We denote $X=\{1,\ldots,m\}$, $Y=\{1,\ldots,n\}$ and let $f:X\times Y \to \{0,1\}$ be the function that defines the computational task by determining the wanted and unwanted input-output pairs $(x,y)\in X\times Y$.
If each $x\in X$ is associated to only one correct answer $g(x)\in Y$, then the wanted pairs constitute the set $\{(x,g(x)):x\in X\}$ and the task is defined by the Kronecker delta function 
$f(x,y) = \delta_{g(x),y}$. More generally, an input $x$ may have several correct answers, which therefore constitute a subset $G_x\subset Y$. In this case, we choose $f(x,y) = 1_{G_x}(y)$, where $1_{G_x}$ is the indicator function of the set $G_x$.

In the current setting, we do not separate the initial quantum state encoding and the quantum process that transforms the states.
This is due to the fact that the anticipative method alters only the final measurement and therefore only the form of the quantum states just before the measurement matters.
We denote $\en(x)=(1/m)\,\Phi(\varrho_x)$, so that the mapping $x \mapsto \en(x)$ describes all what happens to the input $x$ before the quantum measurement is performed. We call $\en$ the {\em state ensemble} of our computational task. The uniform sampling of the inputs is included in the normalization constant $1/m$ of $\en$.
A quantum measurement is mathematically described as a positive operator valued measure $\M$. Denoting by $Z$ the outcome set of $\M$, the probability of choosing the input $x\in X$ and getting the outcome $z\in Z$ is hence $\tr{\en(x) \M(z)}$.

On the classical side, we write $\alpha(S\cond x) = p(S\cond x)$ to denote the conditional probability of getting $S$ from the classical computation when $x\in X$ is the input. We say that $\alpha$ is the {\em partial information map} of our quantum-classical hybrid computation. Note that the formalism allows $S$ can be any information about the task, not just the exclusion of outcomes, but for simplicity we just deal with the exclusion.

Finally, the combination of classical and quantum outcomes is a post-processing that gives the final answer $y\in Y$ with probability $\nu_S(y\cond z) = p(y\cond S,z)$. We refer to $\nu$ as the {\em post-processing map}.
With these notations, the probability of guessing a correct answer is
\begin{equation}
\label{eq:eapostgeneral}
 \Epost = \sum_{x,y,z,S} f(x,y)\,\nu_S(y\cond z)\,\alpha(S\cond x)\,\tr{\en(x) \M(z) } \, .
\end{equation}
We typically assume that the state ensemble $\en$ and the partial information map $\alpha$ are fixed, but we need to find a quantum measurement $\M$ and a post-processing map $\nu$ performed after \cpostt{}, so that the success probability of Eq.~\eqref{eq:eapostgeneral} is as high as possible.

Note, that even though Eq.~\eqref{eq:eapostgeneral} describes the success rate including the arrival of \cpostt{}, its two extreme cases also describe situations when no classical information is provided, or when all wrong answers are excluded. The latter case means that $\alpha(S\cond x) = 1$ if $S=Y\setminus G_x$, where $G_x$ is the set of the correct answers for the input $x$. The optimal post-processing map then necessarily satisfies $\nu_{Y\setminus G_x}(y\cond z)=0$ for all $y\notin G_x$, and the final guess $y$ can be chosen independently of the measurement outcome $z$. In this way, when all wrong answers are excluded, we can trivially achieve the equality $\Epost=1$ for any measurement $\M$. The former case of a situation with no \cpostt{}, instead, corresponds to setting $\alpha(S \cond x)=\alpha(S)$ independently of the input $x$. In this case, the summation over $S$ can be carried out on the post-processings $\nu_S$ and the right-hand side of Eq.~\eqref{eq:eapostgeneral} becomes the probability
\begin{equation}\label{eq:e0general}
\sprob_0 = \sum_{x,y,z} f(x,y)\, \nu_0(y \cond z)\, \tr{\en(x)\M(z)}\,,
\end{equation}
where
\begin{equation}
\nu_0(y\cond z) = \sum_S \nu_S(y\cond z)\,\alpha(S)\,.
\end{equation}

Different optimizations of the previous equations then lead to the cases described above:
 \begin{align}
 \Eapost &= \max_{\M,\nu} \Epost \big(\M,\nu\big)\,,\label{eq:eapost}\\
 \Es_0 &= \,\max_{\M,\nu_0}\, \sprob_0\big(\M,\nu_0\big)\,,\label{eq:es0}
 \end{align}
where in the right-hand side we have explicitely indicated the dependence of the probabilities $\Epost$ and $\sprob_0$ on the chosen measurement $\M$ and the post-processings $\nu$ and $\nu_0$.
If we label the respective optimized measurements as $\Ma$ and $\Ms$, we further have
 \begin{align}
 \Ea_0 &= \,\max_{\nu_0}\, \sprob_0\big(\Ma,\nu_0\big)\,,\label{eq:ea0}\\
 \Espost &= \max_\nu \Epost\big(\Ms,\nu\big)\,.\label{eq:espost}
 \end{align}

%%%%%%%%%%%%%%%%%%%%%%
\subsection{Anticipative measurements with the exclusion of $k$ wrong answers}\label{sec:auxiliary}
%%%%%%%%%%%%%%%%%%%%%%

For simplicity, from now on we assume that the number of wrong answers obtained by means of classical computation is fixed and we denote this number by $k$.
Hence, the set $S$ of the excluded answers is an element of the set
\begin{equation}\label{eq:T}
T = \{ S \subset Y : \mo{S}=k \} \, ,
\end{equation}
which in turn is a subset of the power set $2^Y$. 
We further assume that $S$ is drawn with uniform probability from the subset of all the elements of $T$ which are disjoint from $G_x$.

Having written the computational tasks in Eq.~\eqref{eq:eapostgeneral}, we can interpret it as a quantum guessing game with posterior information and apply the mathematical results developed in \cite{carmeli2018state,carmeli2022quantum}.
The same mathematical formalism can be used to construct and study incompatibility witnesses, but here our aim is different.
One of the main facts contained in the aforementioned works is that any quantum guessing game with posterior information reduces to a usual state discrimination task for an auxiliary state ensemble.
The reduction means that, even if the tasks are different, the optimal measurements are the same and the success probabilities are connected via a simple formula.
To see the reduction in practice, let us first assume that $k=1$, i.e., the classical computation part is giving one wrong answer. Moreover, let us suppose that each input $x$ is associated to only one correct answer $g(x)$.
In this case, we can identify the set $T$ of Eq.~\eqref{eq:T} with $Y$ and write the partial information map as $\alpha(t\cond x) = (1-\delta_{g(x),t})/(n-1)$.
The auxiliary state ensemble, denoted by $\bar{\en}$, is then defined on the Cartesian product $Y^n$ and given as
\begin{equation}
\begin{aligned}
\bar{\en}(y_1,\ldots,y_n) & = \frac{1}{|Y|^{|T|-1}\Delta} \sum_{x,t} f(x,y_t)\,\alpha(t\cond x)\,\en(x) \\
& = \frac{1}{(n-1)\,n^{n-1}} \sum_x r(x)\,\en(x) \,, 
\end{aligned}
\end{equation}
where 
\begin{gather}
\Delta = \sum_{x,y} f(x,y)\,\tr{\en(x)} = 1 \,, \\
r(x) = \big|\{t\in Y : y_t = g(x) \text{ and } t\neq g(x)\}\big|\,.
\end{gather}
With the known techniques of minimum-error state discrimination \cite{Bae13,ha2014discriminating}, one can then find the optimal measurement $\bar{\M}$ that yields the largest success probability in discriminating $\bar{\en}$. 
It turns out that the same measurement $\bar{\M}$ is optimal also for the original guessing game. Indeed, the interpretation of $\bar{\M}$ in the initial setting is that we obtain a measurement outcome that is a tuple, namely $(y_1,\ldots,y_n)\in Y^n$, and the wrong answer given as the posterior information $t$ from the classical computation refines this outcome to the final guess $y_t$.
The success probabilities of the original task and of the auxiliary state discrimination task are not the same but have a simple relation: if the success probability in the auxiliary state discrimination problem is $\Eg$, then
\begin{equation}
\Eapost = n^{n-1}\,\Eg \, .
\end{equation}

The mathematical machinery works in the same way also in the cases with $k>1$. 
Only the form of the auxiliary state ensemble $\bar{\en}$ is different and it can be be obtained from \cite[Sec. 4.4]{carmeli2022quantum}.

%%%%%%%%%%%%%%%%%%%%%%
\section{Qubit application}\label{sec:qubit}
%%%%%%%%%%%%%%%%%%%%%%

While the intention of the anticipative method is to be used for hybrid computation tasks, in order to illustrate the method and test it in current quantum devices we consider a simple example having no direct computational interest, but exhibiting the same quantum features that may appear during quantum computations with a NISQ device. 
In this example, two bits of information are encoded into one qubit, hence the size of the quantum device is smaller than one would need for perfect encoding.
After all quantum processing, we assume that the final states are from two bases.
The task is to infer the values of the original bits with as high success probability as possible.

\begin{figure}
\begin{center}
\includegraphics[height=6cm]{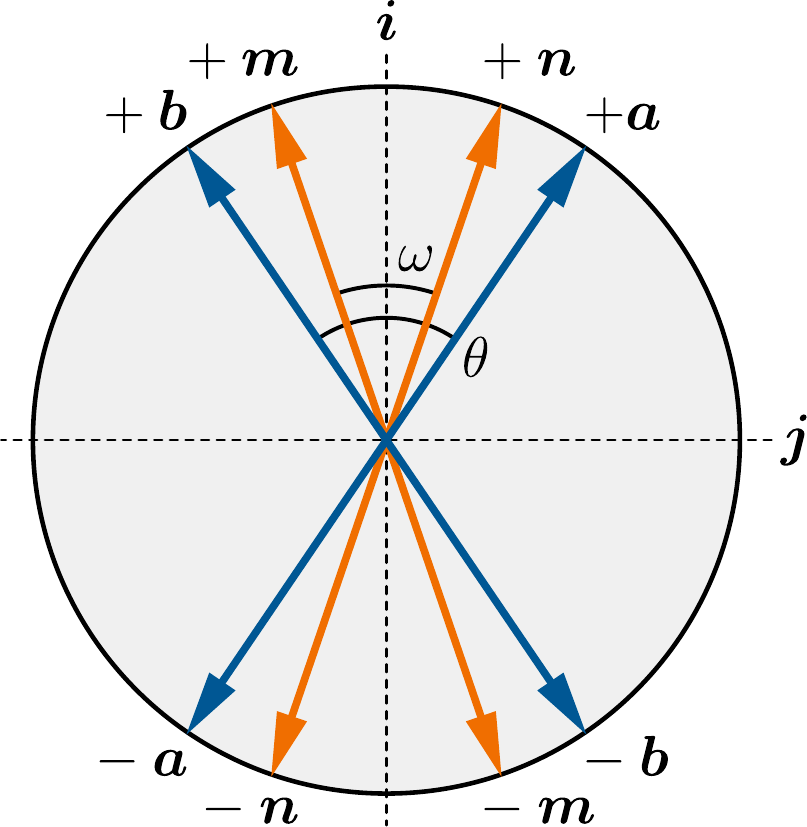}
\end{center}
\caption{Discrimination problem for four qubit states labelled by $\pm \va$ and $\pm \vb$. The standard optimal measurement is a probabilistic projective measurement in the two bases given by the four states. 
The anticipative measurement is a probabilistic projective measurement in the two bases given by the vectors $\pm\vm$ and $\pm\vn$.\label{fig:settings}}
\end{figure}

Qubit states can be identified with vectors on the Bloch sphere.
The final states belong to two bases and the angle between these bases in the Bloch sphere description is parametrized by $\theta\in (0,\pi/2]$.
There is a global unitary freedom to choose the directions, hence we can fix the states corresponding to the Bloch vectors $\pm\va$ and $\pm\vb$ with
\begin{equation}
\begin{aligned}
\va &= \cos\left(\half\theta\right) \vi + \sin\left(\half\theta\right) \vj \,,\\
\vb & = \cos\left(\half\theta\right) \vi - \sin\left(\half\theta\right) \vj \,,
\end{aligned}
\end{equation}
where $\vi$ and $\vj$ are two orthogonal coordinate vectors.
The setting is depicted in Fig.~\ref{fig:settings}.
With respect to the framework explained earlier, we have
\begin{equation}\label{eq:4inputs}
X = \{+\va,\,-\va,\,+\vb,\,-\vb\} 
\end{equation}
and the state ensemble $\en$ is
\begin{equation}\label{eq:4states}
\begin{aligned}
\en(\pm\va) & = \tfrac{1}{8}\left(\id\pm\va\cdot\vsigma\right)\,,\\
\en(\pm\vb) & = \tfrac{1}{8}\left(\id\pm\vb\cdot\vsigma\right)\,.
\end{aligned}
\end{equation}

Although the states within each of the two bases are orthogonal and hence perfectly distinguishable, the four states together are not. 
We will look now on the minimum-error discrimination schemes with or without \cpostt{} both for the standard measurements and the anticipative measurements. 
In the following, we denote by $\Es_k$ and $\Ea_k$ the success probabilities when \cpostt{} consists of $k$ wrong answers. This notation agrees also with the previous case of no \cpostt{} which corresponds to $k=0$. Furthermore, we take $Y=X$ and $f(x,y)=\delta_{x,y}$, which stems from the definition of the problem and our intention to identify the states with highest probability.

%%%%%%%%%%%%%%%%%%%%%%
\subsection{Standard measurement}
%%%%%%%%%%%%%%%%%%%%%%

One of the optimal quantum measurements to distinguish the four different inputs (without \cpostt{}) is
\begin{equation}
\begin{aligned}
\Ms(\pm \va) & = \tfrac{1}{4} \left( \id \pm \va \cdot \vsigma \right)\,,\\
\Ms(\pm \vb) & = \tfrac{1}{4} \left( \id \pm \vb \cdot \vsigma \right)\,.
\end{aligned}
\end{equation}
This measurement results from the maximization of Eq.~\eqref{eq:e0general} with the fixed postprocessing $\nu_0(y\cond z)=\delta_{y,z}$, as can be checked e.g.~by testing the optimality conditions in \cite[Eqs.~(4)-(5)]{Bae13} (the latter choice of $\nu_0$ is not restrictive, since we can always include the post-processing in the optimized measurement).
For the forthcoming calculations we need all the input-output probabilities of the Born rule $\tr{\en(x) \Ms(z)}$, which are given in the following table:\smallskip
\begin{center}
\begin{tabular}{|>{\centering}p{2cm}||>{\centering}p{1.3cm}|>{\centering}p{1.3cm}|>{\centering}p{1.3cm}|>{\centering\arraybackslash}p{1.3cm}|}
\hline
\multirow{2}{*}{\rule{0cm}{0.5cm}\textbf{input} ($x$)} & \multicolumn{4}{c|}{\rule{0cm}{0.4cm} \textbf{measurement outcome} ($z$)} \\[0.1cm]
\cline{2-5}
& \rule{0cm}{0.4cm} $+\va$ & $-\va$ & $+\vb$ & $-\vb$ \\[0.1cm]
\hline\hline
$+\va$ & \rule{0cm}{0.4cm} $\tfrac{1}{8}$ & $0$ & $\tfrac{1}{8} \cos^2\frac{\theta}{2}$ & $\tfrac{1}{8}\sin^2\tfrac{\theta}{2}$ \\[0.1cm] \hline
$-\va$ & \rule{0cm}{0.4cm} $0$ & $\tfrac{1}{8}$  & $\tfrac{1}{8}\sin^2\frac{\theta}{2}$ & $\tfrac{1}{8} \cos^2\frac{\theta}{2}$ \\[0.1cm] \hline
$+\vb$ & \rule{0cm}{0.4cm} $\tfrac{1}{8} \cos^2\frac{\theta}{2}$ &  $\tfrac{1}{8}\sin^2\frac{\theta}{2}$ & $\tfrac{1}{8}$ & $0$ \\[0.1cm] \hline
$-\vb$ & \rule{0cm}{0.4cm} $\tfrac{1}{8}\sin^2\frac{\theta}{2}$ & $\tfrac{1}{8} \cos^2\frac{\theta}{2}$ & $0$ & $\tfrac{1}{8}$ \\[0.1cm]
\hline
\end{tabular}
\end{center}\smallskip
The success probability for the standard measurement without \cpostt{} given by Eq.~\eqref{eq:es0} simply reduces to
\begin{equation}
\Es_0 = \sum_x \tr{\en(x) \Ms(x)} = \frac{1}{2}\,.
\end{equation}
This is the sum of the probabilities from the diagonal of the table.

By using Eq.~\eqref{eq:espost}, we can also infer the success probabilities $\Es_k$ when $k$ outcomes are excluded by \cpostt{}. The optimal strategy consists in guessing the non-excluded input having the highest probability conditioned to the outcome of the measurement $\Ms$.
For $k=1$, $2$ one then obtains the best guess following the mapping rules of the next table:\smallskip
\begin{center}
\begin{tabular}{|>{\centering}p{2.75cm}||>{\centering}p{0.9cm}|>{\centering}p{0.9cm}|>{\centering}p{0.9cm}|>{\centering\arraybackslash}p{0.9cm}|}
\hline
\multirow{2}{*}{\begin{tabular}{c} \rule{0cm}{0.4cm} \textbf{measurement} \\ \textbf{outcome} ($z$) \end{tabular}} & \multicolumn{4}{c|}{\rule{0cm}{0.4cm} \textbf{relabeling priority} ($y$)} \\[0.1cm]
\cline{2-5}
& \rule{0cm}{0.4cm} $1^\mathrm{st}$ & $2^\mathrm{nd}$ & $3^\mathrm{rd}$ & $4^\mathrm{th}$ \\[0.1cm]
\hline\hline
\rule{0cm}{0.4cm} $+\va$ & $+\va$ & $+\vb$ & $-\vb$ & $-\va$ \\[0.1cm] \hline
\rule{0cm}{0.4cm} $-\va$ & $-\va$ & $-\vb$ & $+\vb$ & $+\va$ \\[0.1cm] \hline
\rule{0cm}{0.4cm} $+\vb$ & $+\vb$ & $+\va$ & $-\va$ & $-\vb$ \\[0.1cm] \hline
\rule{0cm}{0.4cm} $-\vb$ & $-\vb$ & $-\va$ & $+\va$ & $+\vb$ \\[0.1cm]
\hline
\end{tabular}
\end{center}\smallskip

This mapping is determined based on the mapping priority where only non-excluded outcomes from the lower right corner of the table are used. For example, having obtained the outcome $+\va$ and excluded the outcomes $\{+\va, -\vb\}$, the next available choice is $+\vb$. If $\{+\va, +\vb\}$ are excluded, the next best choice is $-\vb$.
This observation has also a visual interpretation. 
Looking at the setting from Fig.~\ref{fig:settings}, we see that the mapping priority is ordered from the closest state to the orthogonal one.
Performing the calculations we obtain
\begin{align}
\Es_1 &= \frac{1}{6} \left( 3 + \cos^2\frac{\theta}{2}  \right)\,, \\
\Es_2 & =\frac{1}{6} \left( 4 + \cos^2\frac{\theta}{2}  \right) \, .
\end{align}

\begin{figure*}
\begin{center}
\includegraphics[height=9cm]{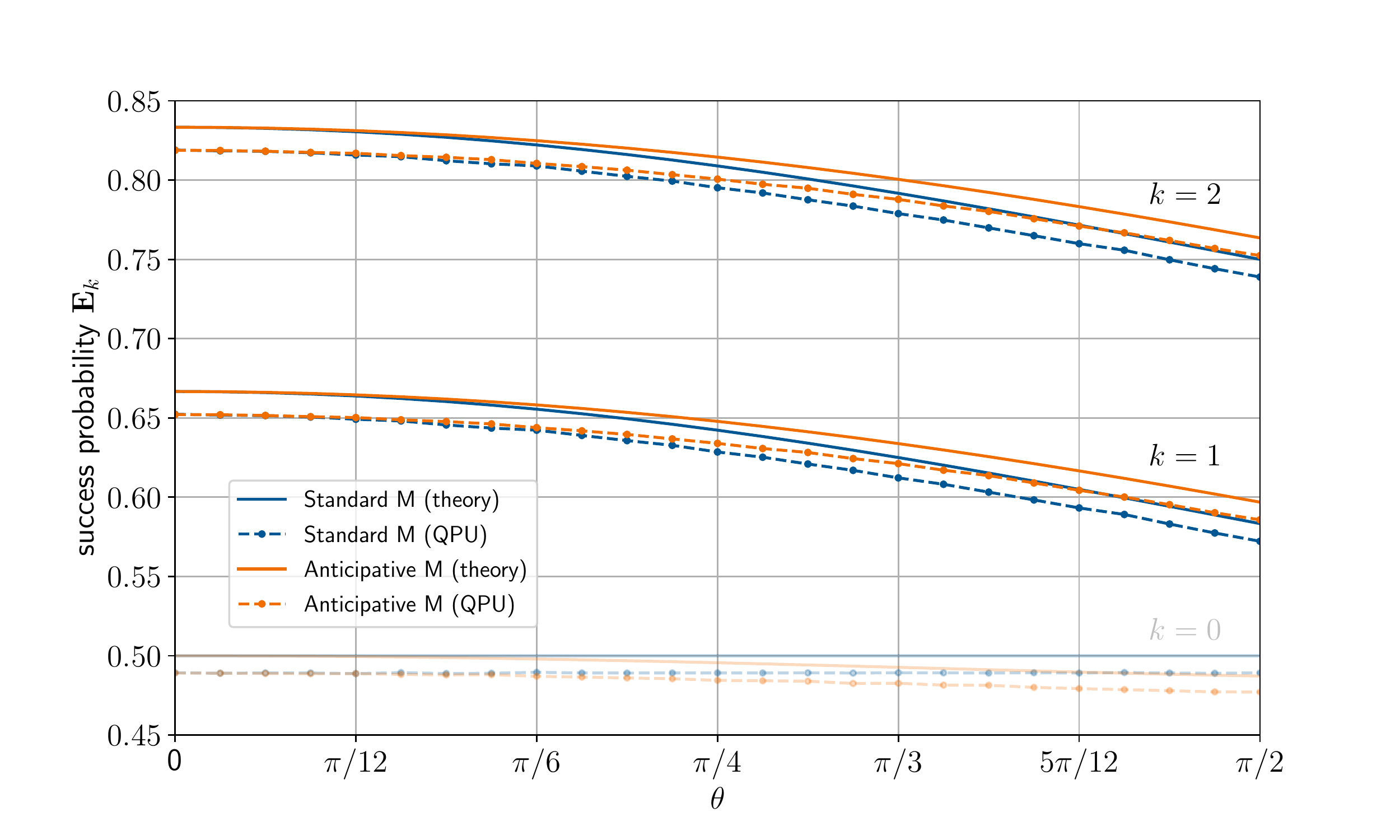}
\end{center}
\caption{Success probabilities of the discrimination task for four qubit states from two bases spanning the angle $\theta$. The comparison is made between standard (blue) and anticipative (orange) measurements for different posterior information --- $k=0$ for measurements without \cpostt{}, $k=1$ for \cpostt{} excluding one wrong answer and $k=2$ for \cpostt{} excluding two wrong answers. Solid lines correspond to the theoretical predictons and dashed lines are results from the \emph{ibmq\_manila}.\label{fig:experiment}}
\end{figure*}

%%%%%%%%%%%%%%%%%%%%%%
\subsection{Anticipative measurement}\label{sec:anti}
%%%%%%%%%%%%%%%%%%%%%%

As explained in Section \ref{sec:anticipativeframework}, the anticipative measurement strategy for the given qubit state ensemble with the exclusion of one or two wrong answers can be calculated with the mathematical framework developed in \cite{carmeli2022quantum}. If there is one excluded answer, then the problem is the same as the quantum guessing game studied in \cite[Sec. 7.4]{carmeli2022quantum} and one of the corresponding anticipative quantum measurement was found to be
\begin{equation}\label{eq:anticipativeM}
\begin{aligned}
\Ma(\pm \vm) & = \tfrac{1}{4} \left( \id \pm \vm \cdot \vsigma \right) \,, \\
\Ma(\pm \vn) & = \tfrac{1}{4} \left( \id \pm \vn \cdot \vsigma \right) \,,
\end{aligned}
\end{equation}
where 
\begin{equation}\label{eq:mn}
\begin{aligned}
\vm & = \frac{\va+3\vb}{\sqrt{10+6\cos\theta}} \,, \\
\vn & = \frac{3\va+\vb}{\sqrt{10+6\cos\theta}} \,.
\end{aligned}
\end{equation}
In Appendix \ref{sec:proof}, we summarize that proof and, in addition, we show that the measurement of Eq.~\eqref{eq:anticipativeM} is optimal also in the case with $k=2$, i.e., when classical information excludes two wrong answers. 
The Born rule probabilities $\tr{\en(x)\Ma(z)}$ are the following:\smallskip
\begin{center}
\begin{tabular}{|>{\centering}p{2cm}||>{\centering}p{1.3cm}|>{\centering}p{1.3cm}|>{\centering}p{1.3cm}|>{\centering\arraybackslash}p{1.3cm}|}
\hline
\multirow{2}{*}{\rule{0cm}{0.5cm}\textbf{input} ($x$)} & \multicolumn{4}{c|}{\rule{0cm}{0.4cm} \textbf{measurement outcome} ($z$)} \\[0.1cm]
\cline{2-5}
& \rule{0cm}{0.4cm} $+\vm$ & $-\vm$ & $+\vn$ & $-\vn$ \\[0.1cm]
\hline\hline
\rule{0cm}{0.4cm} $+\va$ & $Q_+$ & $Q_-$ & $P_+$ & $P_-$ \\[0.1cm] \hline
\rule{0cm}{0.4cm} $-\va$ & $Q_-$ & $Q_+$ & $P_-$ & $P_+$ \\[0.1cm] \hline
\rule{0cm}{0.4cm} $+\vb$ & $P_+$ & $P_-$ & $Q_+$ & $Q_-$ \\[0.1cm] \hline
\rule{0cm}{0.4cm} $-\vb$ & $P_-$ & $P_+$ & $Q_-$ & $Q_+$ \\[0.1cm]
\hline
\end{tabular}
\end{center}\smallskip
Here we have denoted
\begin{equation}
\begin{aligned}
P_\pm &= \frac{1}{16}\left(1\pm\frac{1+3\cos\theta}{\sqrt{10+6\cos\theta}}\right),\\
Q_\pm &= \frac{1}{16}\left(1\pm\frac{\cos\theta + 3}{\sqrt{10+6\cos\theta}}\right).
\end{aligned}
\end{equation}
It is easy to check that $Q_+\geq P_+\geq P_-\geq Q_-$.

Similarly as for the standard measurements, we can visualize the assignment of the measured results to the final guess as a table. Since the input-output probabilities are ordered in the same way as in the standard measurement case and differ only in their magnitude, the mapping follows the same ordering:\smallskip
\begin{center}
\begin{tabular}{|>{\centering}p{2.75cm}||>{\centering}p{0.9cm}|>{\centering}p{0.9cm}|>{\centering}p{0.9cm}|>{\centering\arraybackslash}p{0.9cm}|}
\hline
\multirow{2}{*}{\begin{tabular}{c} \rule{0cm}{0.4cm} \textbf{measurement} \\ \textbf{outcome} ($z$) \end{tabular}} & \multicolumn{4}{c|}{\rule{0cm}{0.4cm} \textbf{relabeling priority} ($y$)} \\[0.1cm]
\cline{2-5}
& \rule{0cm}{0.4cm} $1^\mathrm{st}$ & $2^\mathrm{nd}$ & $3^\mathrm{rd}$ & $4^\mathrm{th}$ \\[0.1cm]
\hline\hline
\rule{0cm}{0.4cm} $+\vn$ & $+\va$ & $+\vb$ & $-\vb$ & $-\va$ \\[0.1cm] \hline
\rule{0cm}{0.4cm} $-\vn$ & $-\va$ & $-\vb$ & $+\vb$ & $+\va$ \\[0.1cm] \hline
\rule{0cm}{0.4cm} $+\vm$ & $+\vb$ & $+\va$ & $-\va$ & $-\vb$ \\[0.1cm] \hline
\rule{0cm}{0.4cm} $-\vm$ & $-\vb$ & $-\va$ & $+\va$ & $+\vb$ \\[0.1cm]
\hline
\end{tabular}
\end{center}\smallskip
This yields the following expressions for the success probabilities defined in Eq.~\eqref{eq:eapost}:
\begin{align}
\Ea_1 & = \frac{1}{12}\left( 4 + \sqrt{10+6\cos\theta} \right),\\
\Ea_2 &= \frac{1}{12}\left( 6 + \sqrt{10+6\cos\theta} \right).
\end{align}

Finally, to calculate the success probability $\Ea_0$ defined in Eq.~\eqref{eq:ea0}, we observe that the optimal assignment without \cpostt{} is $\pm \vm\mapsto\pm \vb$ and $\pm\vn\mapsto\pm\va$, which again consists in guessing the non-excluded input having the highest probability conditioned on the outcome of the measurement $\Ma$. In this way, we get
\begin{equation}
\Ea_0 = 4Q_+ = \frac{1}{4}\left(1 + \frac{\cos\theta + 3}{\sqrt{10+6\cos\theta}}\right)\,.
\end{equation}

All the studied situations are depicted for different $\theta$ angles in Fig.~\ref{fig:experiment} (solid lines). We can notice that while for the case without \cpostt{} ($k=0$) the anticipative measurement operates worse than the standard measurement, with \cpostt{} ($k=1,2$) the anticipative measurement improves the success probability.

\begin{figure*}
\begin{center}
\includegraphics[height=5.5cm]{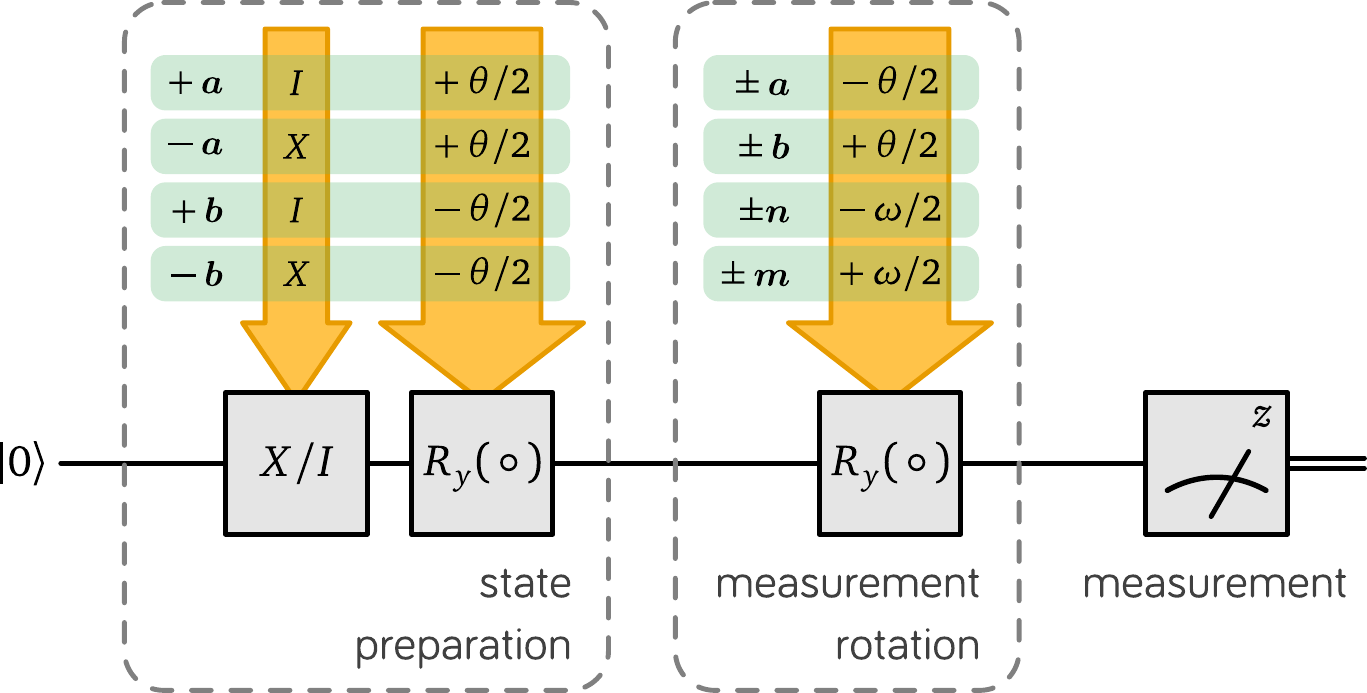}
\end{center}
\caption{Diagram of circuit preparation for the data collection phase. For each choice of the angle $\theta$, the prepared state and the measurement direction, one circuit was constructed and executed $20,\!000$ times.}\label{fig:experiment_setup}
\end{figure*}

%%%%%%%%%%%%%%%%%%%%%%
\subsection{Demonstrations}
%%%%%%%%%%%%%%%%%%%%%%

The investigated state discrimination problem can be easily applied to current quantum devices. 
The possibility to do so comes from the fact that the measurements involved, whether standard or anticipative, are probabilistic projective measurements. These measurements, a subclass of projective simulable measurements \cite{oszmaniec2017simulating,guerini2017operational}, are implementable as a random choice from the set of projective measurements \cite{haapasalo2012quantum}. In the qubit case, this relates to a random choice of a base change prior to the standard $z$-measurement.
For example, the statistics of the measurement $\Ma$ can be obtained as a union of the statistics of the projective measurement in the direction $\pm\vm$ and the projective measurement in the direction $\pm\vn$ with equal number of runs.
If the measurements would not belong to the subclass of projective simulable measurements, we could still implement them but that would require ancillary dimensions. 

Using IBMQ device \emph{ibmq\_manila}, we implemented the measurements in Qiskit, where for the rotations we used only \texttt{RY} gates (more information in Appendix \ref{sec:IBM}). Since the native gate set does not contain the \texttt{RY} gate, prior to the computation it needs to be decomposed using the \texttt{SX} and \texttt{RZ} native gates. This was automatically performed by the transpiler.

For the standard measurement in the basis $\pm \va$ we used the angle $+\theta/2$, while for the measurement in the basis $\pm \vb$ we used the angle $-\theta/2$. For the anticipative measurement we needed to rotate the basis either to the $\pm \vn$ direction by the angle $+\omega/2$ or to the $\pm \vm$ direction by the angle $-\omega/2$. The angle $\omega$ is obtained from the dot product of the vectors $\vm$ and $\vn$ defined in Eq.~\eqref{eq:mn}:
\begin{equation}
\cos\omega = \vm\cdot\vn = \frac{3+5\cos\theta}{5+3\cos\theta}.
\end{equation}
As explained earlier, this angle remains the same in both cases with $k=1$ and $k=2$.

The submissions to \emph{ibmq\_manila} contained circuits with $25$ different choices for the parameter $\theta$, 4 choices of the initial state ($\pm \va$ or $\pm \vb$), 2 choices for the measurement (anticipative / standard), and 2 choices for the measurement basis ($\va$ or $\vb$ for standard and $\vm$ or $\vn$ for anticipative), thus leading to 400 circuits, which were run for $20,\!000$ repetitions each. The preparation scheme of these circuits is illustrated in Fig.~\ref{fig:experiment_setup}.

All computations were performed on qubit 0 with applied optimization for the circuits, possibly combining preparation and measurement rotations into a more convenient form. Collected data were then post-processed to obtain the probabilities. The post-processings were not simulated any more but were computed from obtained data --- each datapoint (executed circuit) was used to obtain the success probability for all allowed exclusions (those not containing the prepared state) and, therefore, it was used multiple times. The number of repetitions was high enough to make variances minimal and so there was no need to keep track of the precise number of shots for each measurement and post-processing, allowing us to extract probabilities this way.

Data obtained in the demonstration is depicted in Fig.~\ref{fig:experiment} as dashed lines. We can observe the general behavior that was theoretically described above --- the anticipative measurement is worse than the standard measurement in the absence of \cpostt{}, but it shows improvement over the standard measurement with \cpostt{} at hand. An interesting point is the region close to $\pi/2$, where we can not only see that the anticipative measurement is better than the standard measurement (for $k=1$ and $k=2$) but, even more, the results of the demonstration show an improvement over the theoretically best standard measurement.

%%%%%%%%%%%%%%%%%%%%%%
\section{Discussion}\label{sec:discussion}
%%%%%%%%%%%%%%%%%%%%%%

In this paper we have investigated a specific type of hybrid quantum-classical computation --- parallel quantum-classical computation where one performs quantum and classical computations simultaneously and combines the results in the end. Quantum computation necessarily ends with a quantum measurement and we have focused on this particular part of the process. We have introduced a method of anticipative quantum measurements, where the quantum measurement is optimized not in isolation but taking into account that there will be classical supplementing information later. 
Crucially, the optimization does not depend on the actual outcome of the classical information but just on the general form of it. 
As an exemplary case we highlighted classical posterior information that rules out some wrong answers.
We demonstrated that the method works even with a noisy real quantum device.

The success of the anticipative method raises some immediate thoughts and questions.
Firstly, the scaling properties of the approach are not studied in this paper, but are important for further considerations. With the scaled dimensions and scaled number of outcomes, we generally observe scaling of the algorithmic time. 
This in turn provides space for additional (time) resources of the parallel classical computation that shall provide a more informative posterior information. It remains for further research to ascertain to what extent a measurable advantage is retained. It might well be that this question is even specific to the chosen task. Still, in the worst case, anticipative measurements might find their place in computations on current NISQ devices with limited resources and small tasks.

Secondly, one could possibly anticipate the classical posterior information not only in the last step of the quantum computation, namely in the quantum measurement, but also in the earlier steps of it.
This would mean that one would modify the quantum process as a whole, allowing for more freedom in the optimization and thus potentially leading even to higher success rates.
A drawback is that the method cannot then anymore be developed in the general form as we have done here, but it should take into account the actual computational task.

Finally, we have observed several peculiar properties in the specific exemplary qubit task. 
Namely, we saw that the anticipative measurement is the same for the cases of excluding one wrong answer and two wrong answers although in general every type of \cpostt{} can lead to a different solution. 
Further, in these cases the anticipative measurement has only four outcomes that are nonzero. 
As explained in Section \ref{sec:auxiliary} the anticipative measurement is optimal also with respect to the auxiliary state ensemble discrimination, hence in both cases one would have expected more outcomes than four.  
A further investigation will show to what extent these observations are general and to what extent they are specific to the task.

%%%%%%%%%%%%%%%%%%%%%%
\section{Acknowledgment}
%%%%%%%%%%%%%%%%%%%%%%

We acknowledge the use of IBM Quantum services for this work. The views expressed are those of the authors, and do not reflect the official policy or position of IBM or the IBM Quantum team.

\newpage

%%%%%%%%%%%%%%%%%%%%%%
\appendix
%%%%%%%%%%%%%%%%%%%%%%

%%%%%%%%%%%%%%%%%%%%%%
\section{Anticipative qubit measurements}\label{sec:proof}
%%%%%%%%%%%%%%%%%%%%%%

We fix the set of inputs $X$ and the state ensemble $\en$ as in Eqs.~\eqref{eq:4inputs} and \eqref{eq:4states}, respectively. Our task is to discriminate any given input $\vx\in X$, so we let the output and input sets coincide, $Y=X$, and we choose $f(\vx,\vy) = \delta_{\vx,\vy}$ following the notations of Section \ref{sec:anticipativeframework} (we use the bold symbols $\vx$, $\vy$ in place of $x$, $y$ to stress that we deal with vectors in $\R^3$). We consider the parallel quantum-classical computation in which CPOST consists in excluding $k$ wrong answers with either $k=1$ or $k=2$. The set $S$ of the excluded answers is thus an element of the collection of sets $T$ defined in Eq.~\eqref{eq:T}. More precisely, for any fixed value of the input $\vx$ the set $S$ is drawn from the subcollection
\begin{equation}
T_\vx = \{ S \subset Y : \mo{S}=k \text{ and } \vx\notin S \}\,.
\end{equation}
We further assume that $S$ is randomly picked with uniform probability within $T_\vx$, which means that the partial information map $\alpha$ is
\begin{equation}
\alpha(S\cond\vx) = \tfrac{1}{3} \,1_{X\setminus S}(\vx) \quad \forall S\in T\,.
\end{equation}
In the particular case with $k=1$, we can canonically identify $T\simeq X$ and $T_\vx\simeq X\setminus\{\vx\}$, so that the partial information map is $\alpha(\vt\cond\vx) = (1-\delta_{\vx,\vt})/(m-1)$ as described in Section \ref{sec:auxiliary}.

If we perform a measurement $\M$ with the outcome set $Z$ and we post-process its outcome by means of a post-processing map $\nu$, we guess a correct answer with the probability $\Epost$ of Eq.~\eqref{eq:eapostgeneral}. As we did in Eqs.~\eqref{eq:eapost} and \eqref{eq:espost}, we rewrite such probability as $\Epost(\M,\nu)$ in order to stress its dependence on $\M$ and $\nu$. Further, following Eqs.~\eqref{eq:eapost}, we denote by $\Eapost$ the value of $\Epost(\M,\nu)$ optimized over all measurements $\M$ and all post-processing maps $\nu$. The aim of this appendix is then to determine the maximum $\Eapost$, find the measurements $\Ma$ at which it is attained and describe the corresponding optimal post-processings. The essential tool is the following relation proved in \cite[Prop.~2]{carmeli2022quantum}:
\begin{equation}\label{eq:max_barM}
\Eapost = \max_{\bar{\M}} \Epost(\bar{\M},\pi) \,.
\end{equation}
In this expression, the maximum on the right-hand side ranges over all measurements $\bar{\M}$ with the outcome set $X^T$, where $X^T$ is the set of all functions $\phi:T\to X$. Moreover, $\pi$ denotes the post-processing map
\begin{equation}
\pi_S(\vx\cond\phi) = \delta_{\vx,\phi(S)} \,.
\end{equation}
Note that the outcome set $X^T$ of $\bar{\M}$ and the post-processing map $\pi$ are fixed in the right-hand side of Eq.~\eqref{eq:max_barM}. This compares with the definition of $\Eapost$, which in principle would require to optimize $\Epost$ over both $\M$ and $\nu$, also allowing the outcome set of $\M$ to vary among all finite sets.

\begin{theorem}\label{thm:ecluding_spin}
For $j\in\{+,-\}$, let $\phi^{\va,\vb}_j$ be element of $X^T$ defined as follows:
\begin{subequations}\label{eq:phiabj}
\begin{enumerate}[-]
\item if $k=1$,
\begin{equation}
\phi^{\va,\vb}_j(S) = \begin{cases}
j\va & \text{ if } j\va\notin S\\
j\vb & \text{ if } j\va\in S
\end{cases}\,,
\end{equation}
\item if $k=2$,
\begin{equation}
\phi^{\va,\vb}_j(S) = \begin{cases}
j\va & \text{ if } j\va\notin S\\
j\vb & \text{ if } j\va\in S \text{ and } j\vb\notin S\\
-j\vb & \text{ if } j\va\in S \text{ and } j\vb\in S
\end{cases}\,.
\end{equation}
\end{enumerate}
\end{subequations}
Then, a measurement attaining the maximum in Eq.~\eqref{eq:max_barM} is the measurement $\bar{\M} = \bar{\M}^{\va,\vb}$ given by
\begin{equation}\label{eq:barMab}
\begin{aligned}
\bar{\M}^{\va,\vb}(\phi) & = 0 \quad \text{for $\phi\notin\{\phi^{\va,\vb}_+,\phi^{\va,\vb}_-\}$}\,,\\
\bar{\M}^{\va,\vb}(\phi^{\va,\vb}_j) & = \tfrac{1}{2}\left(\id + j\,\frac{3\va+\vb}{\no{3\va+\vb}}\cdot\vsigma\right)\,.
\end{aligned}
\end{equation}
Moreover, we have
\begin{equation*}
\Eapost = \frac{1}{12}\left( 2 + 2r + \sqrt{10+6\,\va\cdot\vb}\right).
\end{equation*}
\end{theorem}
By exchanging the vectors $\va$ and $\vb$ in Eqs.~\eqref{eq:phiabj} and \eqref{eq:barMab}, we also obtain the optimal measurement
\begin{equation}\label{eq:barMba}
\begin{aligned}
\bar{\M}^{\vb,\va}(\phi) & = 0 \quad \text{for $\phi\notin\{\phi^{\vb,\va}_+,\phi^{\vb,\va}_-\}$}\,, \\
\bar{\M}^{\vb,\va}(\phi^{\vb,\va}_j) & = \frac{1}{2}\left(\id + j\,\frac{\va+3\vb}{\no{\va+3\vb}}\cdot\vsigma\right)\,,
\end{aligned}
\end{equation}
and any convex combination of $\bar{\M}^{\va,\vb}$ and $\bar{\M}^{\vb,\va}$ is still a measurement attaining the maximum in Eq.~\eqref{eq:max_barM}.
As an example, the optimal measurement for the case with $k=1$ that was derived in \cite[Equations (65) and (66)]{carmeli2022quantum} coincides with the sum $(1/2)\,\bar{\M}^{\va,\vb} + (1/2)\,\bar{\M}^{\vb,\va}$. In the particular case with $\va\cdot\vb =0$, Theorem \ref{thm:ecluding_spin} yields two additional optimal measurements besides $\bar{\M}^{\va,\vb}$ and $\bar{\M}^{\vb,\va}$, namely the measurements $\bar{\M}^{-\va,\vb}$ and $\bar{\M}^{\vb,-\va}$ that are obtained by replacing $\va$ with $-\va$ in \eqref{eq:barMab} and \eqref{eq:barMba}, respectively. The supports of the four measurements $\bar{\M}^{\va,\vb}$, $\bar{\M}^{\vb,\va}$, $\bar{\M}^{-\va,\vb}$ and $\bar{\M}^{\vb,-\va}$ are mutually disjoint subsets of $X^T$, and each of them contains two elements. By restricting the convex sum $(1/2)\,\bar{\M}^{\va,\vb} + (1/2)\,\bar{\M}^{\vb,\va}$ to its support $\{\phi^{\va,\vb}_+,\phi^{\va,\vb}_-,\phi^{\vb,\va}_+,\phi^{\vb,\va}_-\}$, evaluating the corresponding restriction of the post-processing map $\pi$ and relabeling $\phi^{\va,\vb}_j\to j\vn$ and $\phi^{\vb,\va}_j\to j\vm$, we reduce the sum $(1/2)\,\bar{\M}^{\va,\vb} + (1/2)\,\bar{\M}^{\vb,\va}$ to the anticipative measurement $\Ma$ of Eq.~\eqref{eq:anticipativeM}, while $\pi$ becomes the post-processing map $\nu$ with
\begin{enumerate}[-]
\item if $k=1$,
\begin{align*}
\nu_S(\vx\cond j\vn) & = \delta_{\vx,j\va}\,1_{X\setminus S}(j\va) + \delta_{\vx,j\vb}\,1_S(j\va)\,, \\
\nu_S(\vx\cond j\vm) & = \delta_{\vx,j\vb}\,1_{X\setminus S}(j\vb) + \delta_{\vx,j\va}\,1_S(j\vb)\,,
\end{align*}
\item if $k=2$,
\begin{align*}
\nu_S(\vx\cond j\vn) & = \delta_{\vx,j\va}\,1_{X\setminus S}(j\va) + \delta_{\vx,j\vb}\,1_S(j\va)\,1_{X\setminus S}(j\vb) \\
& \quad + \delta_{\vx,-j\vb}\,1_S(j\va)\,1_S(j\vb)\,, \\
\nu_S(\vx\cond j\vm) & = \delta_{\vx,j\vb}\,1_{X\setminus S}(j\vb) + \delta_{\vx,j\va}\,1_{X\setminus S}(j\va)\,1_S(j\vb) \\
& \quad + \delta_{\vx,-j\va}\,1_S(j\va)\,1_S(j\vb)\,.
\end{align*}
\end{enumerate}
We observe that $\nu$ coincides with the relabeling priority described in the second table of Section \ref{sec:anti} for both cases with $k=1$ and $k=2$.

\begin{proof}[Proof of Theorem \ref{thm:ecluding_spin}]
By \cite[Eq.~(25)]{carmeli2022quantum}, we have
$$
\Eg(\bar{\M},\pi) = C\,\Eg_{\bar{\en}}(\bar{\M}) \,,
$$
where $\bar{\en}:X^T\to\lc$ is an auxiliary state ensemble associated with $\en$ and defined as
$$
\bar{\en}(\phi) = \frac{1}{24\,C}\sum_\vx \big|\phi^{-1}(\vx)\cap T_\vx\big|\left(\id+\vx\cdot\vsigma\right) \,,
$$
the scalar $C>0$ is a suitable normalization constant, and
$$
\Eg_{\bar{\en}}(\bar{\M}) = \sum_\phi \tr{\bar{\en}(\phi)\bar{\M}(\phi)}
$$
is the success probability of the usual state discrimination task for the auxiliary state ensemble $\bar{\en}$ and the measurement $\bar{\M}$. In particular, the measurements $\bar{\M}$ which maximize the probabilities $\Epost(\bar{\M},\pi)$ and $\Eg_{\bar{\en}}(\bar{\M})$ coincide.
Thus, optimizing the state discrimination task with posterior information for the original ensemble $\en$ reduces to a standard state discrimination problem for the auxiliary ensemble $\bar{\en}$. To solve the latter problem, we denote by $\Lambda(\bar{\en})$ the largest eigenvalue of all the operators $\bar{\en}(\phi)$, $\phi\in X^T$, and we observe that any measurement $\bar{\M}$ satisfies the inequality
\begin{equation*}
\Eg_{\bar{\en}}(\bar{\M}) \leq 2 \, \Lambda(\bar{\en})\,.
\end{equation*}
As a consequence of \cite[Proposition 1]{carmeli2022quantum}, the above bound is attained if and only if $\bar{\M}$ satisfies the relation 
\begin{equation*}
\bar{\en}(\phi)\,\bar{\M}(\phi) = \Lambda(\bar{\en})\,\bar{\M}(\phi)
\end{equation*}
for all $\phi\in X^T$. 
We will show that this is actually the case for the measurement $\bar{\M}^{\va,\vb}$ of \eqref{eq:barMab}, and therefore
\begin{equation*}
\Epost(\bar{\M}^{\va,\vb},\pi) = \max_{\bar{\M}} \Epost(\bar{\M},\pi) = 2\,C\, \Lambda(\bar{\en})
\end{equation*}
for the value of $\Lambda(\bar{\en})$ that we will explicitly determine below.\\
For notational convenience, for all $\phi\in X^T$ and $j\in\{+,-\}$, we introduce the nonnegative integer numbers
\begin{equation*}
\alpha^\phi_j = \mo{\phi^{-1}(j\va)\cap T_{j\va}}\,,\qquad\quad \beta^\phi_j = \big|\phi^{-1}(j\vb)\cap T_{j\vb}\big|\,,
\end{equation*}
so that the auxiliary state ensemble rewrites
\begin{align*}
\bar{\en}(\phi) = \,& \frac{1}{24\,C}\, \Big\{ \big(\alpha^\phi_+ + \alpha^\phi_- + \beta^\phi_+ + \beta^\phi_-\big)\,\id \\
& + \Big[\big(\alpha^\phi_+ - \alpha^\phi_-\big)\,\va + \big(\beta^\phi_+ - \beta^\phi_-\big)\,\vb\Big]\cdot\vsigma\Big\}\,.
\end{align*}
The largest eigenvalue of the operator $\bar{\en}(\phi)$ is
\begin{equation*}
\begin{aligned}
\lambda(\phi) = \,& \frac{1}{24\,C}\, \Big\{\alpha^\phi_+ + \alpha^\phi_- + \beta^\phi_+ + \beta^\phi_- \\
& + \Big\|\big(\alpha^\phi_+ - \alpha^\phi_-\big)\,\va + \big(\beta^\phi_+ - \beta^\phi_-\big)\,\vb\Big\|\Big\} \\
= & \frac{1}{24\,C}\,\gamma\big(\alpha^\phi_+,\,\alpha^\phi_-,\,\beta^\phi_+,\,\beta^\phi_-\big)\,,
\end{aligned}
\end{equation*}
where $\gamma$ is the function
\begin{equation*}
\begin{aligned}
& \gamma\big(\alpha_+,\,\alpha_-,\,\beta_+,\,\beta_-\big) = \alpha_+ + \alpha_- + \beta_+ + \beta_- \\
& \qquad + \big[\big(\alpha_+ - \alpha_-\big)^2 + \big(\beta_+ - \beta_-\big)^2 \\
& \qquad + 2\,\big(\alpha_+ - \alpha_-\big)\big(\beta_+ - \beta_-\big)\,\va\cdot\vb\big]^{\frac{1}{2}} \,.
\end{aligned}
\end{equation*}
If $\alpha^\phi_+\neq\alpha^\phi_-$ or $\beta^\phi_+\neq\beta^\phi_-$, the projection onto the $\lambda(\phi)$-eigenspace of $\bar{\en}(\phi)$ is the rank-$1$ operator
\begin{equation*}
\Pi(\phi) = \frac{1}{2}\left(\id + \frac{\big(\alpha^\phi_+ - \alpha^\phi_-\big)\,\va + \big(\beta^\phi_+ - \beta^\phi_-\big)\,\vb}{\no{\big(\alpha^\phi_+ - \alpha^\phi_-\big)\,\va + \big(\beta^\phi_+ - \beta^\phi_-\big)\,\vb}}\cdot\vsigma\right)\,.
\end{equation*}
We now proceed to separately evaluate
$$
\Lambda(\bar{\en}) = \max_\phi \lambda(\phi) = \frac{1}{24\,C}\,\max_\phi \gamma\big(\alpha^\phi_+,\,\alpha^\phi_-,\,\beta^\phi_+,\,\beta^\phi_-\big)
$$
in the two cases with $k=1$ and $k=2$.

\begin{enumerate}[-]
\item {\em Case $k=1$.} Since $|T_\vx| = 3$ for all $\vx\in X$ and $\sum_\vx \big|\phi^{-1}(\vx)\big| = |T| = 4$, the numbers $\alpha^\phi_+$, $\alpha^\phi_-$, $\beta^\phi_+$ and $\beta^\phi_-$ satisfy the constraints
\begin{align*}
& \alpha^\phi_+,\,\alpha^\phi_-,\,\beta^\phi_+,\,\beta^\phi_- \in\{0,1,2,3\}\,, \\
& \alpha^\phi_+ + \alpha^\phi_- + \beta^\phi_+ + \beta^\phi_- \leq 4\,.
\end{align*}
For $\va\cdot\vb>0$, the constrained maximum
\begin{equation}\tag{$\ast$}\label{eq:constr_max_1}
\begin{aligned}
& \max\gamma\big(\alpha_+,\,\alpha_-,\,\beta_+,\,\beta_-\big)\quad\text{subject to} \\
& \qquad\qquad\qquad\quad \alpha_+,\,\alpha_-,\,\beta_+,\,\beta_- \in\{0,1,2,3\}\,, \\
& \qquad\qquad\qquad\quad \alpha_+ + \alpha_- + \beta_+ + \beta_- \leq 4
\end{aligned}
\end{equation}
was evaluated in \cite[Appendix A]{carmeli2022quantum} and found to be equal to $4+\sqrt{10+6\,\va\cdot\vb}$. By an easy continuity argument, this result extends also to the case with $\va\cdot\vb = 0$. We have
$$
\big(\alpha^\phi_+,\,\alpha^\phi_-,\,\beta^\phi_+,\,\beta^\phi_-\big) = \begin{cases}
(3,0,1,0) & \text{if } \phi = \phi^{\va,\vb}_+ \\
(0,3,0,1) & \text{if } \phi = \phi^{\va,\vb}_-
\end{cases}\,,
$$
which are two feasible points attaining the maximum \eqref{eq:constr_max_1}. Therefore,
$$
\Lambda(\bar{\en}) = \frac{1}{24\,C}\left(4+\sqrt{10+6\,\va\cdot\vb}\right) \,.
$$
\item {\em Case $k=2$.} We still have $|T_\vx| = 3$ for all $\vx\in X$, but now $\sum_\vx \big|\phi^{-1}(\vx)\big| = |T| = 6$. Moreover, if $\vx\neq\vy$, we have $\{\vx,\vy\}\notin T_\vx\cup T_\vy$ and $\phi^{-1}(\vx)\cap\phi^{-1}(\vy)=\emptyset$, hence it must be $\big|\phi^{-1}(\vx)\cap T_\vx\big|+\big|\phi^{-1}(\vy)\cap T_\vy\big|\leq |T|-1 = 5$. It follows that the numbers $\alpha^\phi_+$, $\alpha^\phi_-$, $\beta^\phi_+$ and $\beta^\phi_-$ now satisfy the constraints
\begin{align*}
& \alpha^\phi_+,\,\alpha^\phi_-,\,\beta^\phi_+,\,\beta^\phi_- \in\{0,1,2,3\}\,, \\
& \alpha^\phi_+ + \alpha^\phi_- + \beta^\phi_+ + \beta^\phi_- \leq 6\,,\\
& \alpha^\phi_+ + \beta^\phi_+\leq 5\,, \\
& \alpha^\phi_- + \beta^\phi_-\leq 5\,.
\end{align*}
As in the previous case, we first evaluate the constrained maximum
\begin{equation}\tag{$\ast\ast$}\label{eq:constr_max_2}
\begin{aligned}
& \max\gamma\big(\alpha_+,\,\alpha_-,\,\beta_+,\,\beta_-\big)\quad\text{subject to} \\
& \qquad\qquad\qquad\quad \alpha_+,\,\alpha_-,\,\beta_+,\,\beta_- \in\{0,1,2,3\}\,,\\
& \qquad\qquad\qquad\quad \alpha_+ + \alpha_- + \beta_+ + \beta_- \leq 6\,,\\
& \qquad\qquad\qquad\quad \alpha_+ + \beta_+\leq 5\,, \\
& \qquad\qquad\qquad\quad \alpha_- + \beta_-\leq 5\,,
\end{aligned}
\end{equation}
and next we prove that $\big(\alpha^\phi_+,\,\alpha^\phi_-,\,\beta^\phi_+,\,\beta^\phi_-\big)$ is an optimal point for all $\phi\in\{\phi^{\va,\vb}_+,\phi^{\va,\vb}_-\}$. To this aim, we fix a feasible point $\big(\alpha_+,\,\alpha_-,\,\beta_+,\,\beta_-\big)$ and we start by assuming that $\alpha_j - \alpha_h = 3$ for some $j,h\in\{+,-\}$. It follows that $\alpha_j = 3$, $\alpha_h = 0$ and
\begin{align*}
& \gamma\big(\alpha_+,\,\alpha_-,\,\beta_+,\,\beta_-\big) = 3 + \beta_+ + \beta_- \\
& \qquad\qquad + \big[9+\big(\beta_+ - \beta_-\big)^2+6\,\big(\beta_j - \beta_h\big)\,\va\cdot\vb\big]^\frac{1}{2}
\end{align*}
because $\alpha_+,\,\alpha_-\in\{0,1,2,3\}$. If $(\alpha_+,\,\alpha_-,\,\beta_+,\,\beta_-\big)$ is a feasible point with $\beta_j \leq \beta_k$, then $(\alpha'_+,\,\alpha'_-,\,\beta'_+,\,\beta'_-\big) = (\alpha_+,\,\alpha_-,\,\beta_-,\,\beta_+\big)$ is a feasible point such that $\gamma(\alpha'_+,\,\alpha'_-,\,\beta'_+,\,\beta'_-\big) \geq \gamma(\alpha_+,\,\alpha_-,\,\beta_+,\,\beta_-\big)$ and $\beta'_j \geq \beta'_k$. Therefore, in order to find the constrained maximum \eqref{eq:constr_max_2}, we can restrict to the feasible points which satisfy $\beta_j \geq \beta_k$. For any such point, the relation $\alpha_j + \beta_j\leq 5$ requires that $\beta_j\leq 2$. The possibility $\beta_+ = \beta_- = 2$ is excluded by the inequality $\alpha_+ + \alpha_- + \beta_+ + \beta_- \leq 6$. Therefore, the only remaining possibilities are
\begin{itemize}
\item $\big(\beta_j,\,\beta_k) = (2,1)$, and then
$$
\gamma\big(\alpha_+,\,\alpha_-,\,\beta_+,\,\beta_-\big) = 6+\sqrt{10+6\,\va\cdot\vb}\,;
$$
\item $\big(\beta_j,\,\beta_k) = (2,0)$, and then
\begin{align*}
\gamma\big(\alpha_+,\,\alpha_-,\,\beta_+,\,\beta_-\big) & = 5+\sqrt{13+12\,\va\cdot\vb} \\
& \leq 6+\sqrt{10+6\,\va\cdot\vb}\,;
\end{align*}
\item $\big(\beta_j,\,\beta_k) = (1,0)$,  and then
\begin{align*}
\gamma\big(\alpha_+,\,\alpha_-,\,\beta_+,\,\beta_-\big) & = 4+\sqrt{10+6\,\va\cdot\vb} \\
& < 6+\sqrt{10+6\,\va\cdot\vb}\,;
\end{align*}
\item $\big(\beta_j,\,\beta_k) = (1,1)$ or $\big(\beta_j,\,\beta_k) = (0,0)$, and then
\begin{align*}
\gamma\big(\alpha_+,\,\alpha_-,\,\beta_+,\,\beta_-\big) & = 6 + \beta_+ + \beta_- \leq 8 \\
& < 6+\sqrt{10+6\,\va\cdot\vb}\,.
\end{align*}
\end{itemize}
Next, we consider the case with $\beta_j - \beta_h = 3$, and we observe that it is similar to the previous case with $\alpha_j - \alpha_h = 3$. In particular,
$$
\gamma\big(\alpha_+,\,\alpha_-,\,\beta_+,\,\beta_-\big) \leq 6+\sqrt{10+6\,\va\cdot\vb}\,;
$$
also in this case. Finally, since the constraint $\alpha_+,\,\alpha_-,\,\beta_+,\,\beta_-\in\{0,1,2,3\}$ requires that $\max\{|\alpha_+ - \alpha_-|,|\beta_+ - \beta_-|\}\leq 3$, the only remaining possibility is $\max\{|\alpha_+ - \alpha_-|,|\beta_+ - \beta_-|\}\leq 2$, which implies
\begin{align*}
\gamma\big(\alpha_+,\,\alpha_-,\,\beta_+,\,\beta_-\big) & \leq 6+\sqrt{8+8\,\va\cdot\vb} \\
& \leq 6+\sqrt{10+6\,\va\cdot\vb}\,.
\end{align*}
In summary, the constrained maximum \eqref{eq:constr_max_2} is $6+\sqrt{10+6\,\va\cdot\vb}$, and it is attained e.g.~at the feasible points
$$
\big(\alpha^\phi_+,\,\alpha^\phi_-,\,\beta^\phi_+,\,\beta^\phi_-\big) = \begin{cases}
(3,0,2,1) & \text{if } \phi = \phi^{\va,\vb}_+ \\
(0,3,1,2) & \text{if } \phi = \phi^{\va,\vb}_-
\end{cases}\,.
$$
Similarly to the case with $k=1$, it then follows that
$$
\Lambda(\bar{\en}) = \frac{1}{24\,C}\left(6+\sqrt{10+6\,\va\cdot\vb}\right) \,.
$$
\end{enumerate}
In both cases with $k=1$ and $k=2$, the measurement \eqref{eq:barMab} satisfies the equality
$$
\bar{\M}^{\va,\vb}(\phi) = \begin{cases}
\Pi(\phi) & \text{if } \phi \in\{\phi^{\va,\vb}_+,\phi^{\va,\vb}_-\} \\
0 & \text{otherwise}
\end{cases}\,,
$$
hence $\bar{\en}(\phi)\,\bar{\M}^{\va,\vb}(\phi) = \Lambda(\bar{\en})\,\bar{\M}^{\va,\vb}(\phi)$ for all $\phi\in X^T$. The relation \eqref{eq:max_barM} and the optimality of $\bar{\M}^{\va,\vb}$ then follow from the previous discussion.
\end{proof}

%%%%%%%%%%%%%%%%%%%%%%
\section{IBMQ Manila specifications}
%%%%%%%%%%%%%%%%%%%%%%
\label{sec:IBM}

Our demonstrations were performed on \emph{ibmq\_manila,} which is one of the 5-qubit IBM Quantum Falcon Processors, r5.11L (linearly coupled qubits), at that time having backend version 1.0.17 (dated 9.~November 2021). Even though we performed computations on qubit 0 only, we present data for all qubits in the table, where we list qubit frequencies $f$, $T_1$ and $T_2$ times, single qubit errors $\epsilon_1$ and $z$-measurement errors $\epsilon_M$.

The native gate set is \texttt{CX} (controlled \texttt{X}), \texttt{ID} (identity), \texttt{IF\_ELSE} (classical dynamical coditioning), \texttt{RZ} (parametric $z$-rotation), \texttt{SX} (square root of \texttt{X}), \texttt{X} (bit flip / \texttt{NOT} operation in the computational $z$-basis).

For completeness we describe also some of the operations. The gate \texttt{X} can be represented by the Pauli $\sigma_x$ matrix and as a result \texttt{SX} performs a unitary rotation described by the matrix
\[
\sqrt{\sigma_x} = \frac{1}{2}\begin{pmatrix} 1+i & 1-i \\ 1-i & 1+i\end{pmatrix}.
\]
The gate \texttt{RZ} performs a parametric unitary rotation given by the matrix
\[
R_z(\theta) = \begin{pmatrix} e^{-i\frac{\theta}{2}} & 0 \\ 0 & e^{+i\frac{\theta}{2}} \end{pmatrix}.
\]
Finally, we have used the gate \texttt{RY} which performs the parametric rotation
\[
R_y(\theta) = \begin{pmatrix} \cos\frac{\theta}{2} & -\sin\frac{\theta}{2} \\ \sin\frac{\theta}{2} & \cos\frac{\theta}{2}\end{pmatrix}.
\]
Since this gate does not belong to the native gate set of \emph{ibmq\_manila}, it needs to be further decomposed. One can use the identity
\[
R_y(\theta) = i\sqrt{\sigma_x} R_z(\pi-\theta)\sqrt{\sigma_x} R_z(\pi).
\]
As we used automatic regime of Qiskit transpiler, the actual decomposition might vary and is possibly further optimized by the transpiler together with the rest of the optimized circuit.

\smallskip
\begin{center}
\begin{tabular}{|c||c|c|c|c|c|}
\hline
qubit & $f$ [GHz] & $T_1$ [$\mu$s] & $T_2$ [$\mu$s] & $\epsilon_1$ & $\epsilon_M$\\
\hline\hline
0 & $4.96(3)$ & 208 & 116 & $2.06\times 10^{-4}$ & $2.30\times 10^{-2}$\\
1 & $4.83(8)$ & 227 & 86 & $2.46\times 10^{-4}$ & $2.80\times 10^{-2}$\\
2 & $5.03(7)$ & 179 & 25 & $2.41\times 10^{-4}$ & $2.27\times 10^{-2}$\\ 
3 & $4.95(1)$ & 134 & 62 & $1.69\times 10^{-4}$ & $2.14\times 10^{-2}$\\ 
4 & $5.06(6)$ & 147 & 41 & $3.13\times 10^{-4}$ & $2.29\times 10^{-2}$\\ 
\hline
\end{tabular}
\end{center}

%%%%%%%%%
%%%%%%%%%
\end{document}